\documentclass[%
 reprint,
 superscriptaddress,preprintnumbers,
 nofootinbib,
 amssymb,
 aps,
]{revtex4-1}

\usepackage[utf8]{inputenc}
\usepackage{hyperref}
\usepackage{amsmath,amssymb,mathtools}
\usepackage{dsfont}
\usepackage{graphicx}   
\usepackage[table]{xcolor}
\usepackage{colortbl}
\usepackage{url}
\usepackage[normalem]{ulem}
\usepackage{slashed}
\usepackage[capitalise, english]{cleveref}
\usepackage[noindentafter]{titlesec} 
\allowdisplaybreaks

\usepackage{siunitx}
\usepackage{xspace}

    \makeatletter
    \def\CT@@do@color{%
      \global\let\CT@do@color\relax
            \@tempdima\wd\z@
            \advance\@tempdima\@tempdimb
            \advance\@tempdima\@tempdimc
    \advance\@tempdimb\tabcolsep
    \advance\@tempdimc\tabcolsep
    \advance\@tempdima2\tabcolsep
            \kern-\@tempdimb
            \leaders\vrule
                    \hskip\@tempdima\@plus  1fill
            \kern-\@tempdimc
            \hskip-\wd\z@ \@plus -1fill }
    \makeatother

\usepackage{accents}
\DeclareMathSymbol{\widetildesym}{\mathord}{largesymbols}{"65}

\def\thesubsection{\arabic{section}.\arabic{subsection}}

\def\thesection{\arabic{section}}
\titleformat*{\subsubsection}{\normalfont \small \bfseries \boldmath}
\renewcommand{\paragraph}[1]{\vspace{.3em} \indent {\bfseries \boldmath #1 ---}\xspace }
\makeatletter
    \renewcommand{\p@subsection}{}
    \renewcommand{\p@subsubsection}{}
\makeatother

\definecolor{red}{rgb}{0.6,.0706,.1373}
\definecolor{blue}{rgb}{0,0.396,0.741}
\newcommand\myshade{80}
\colorlet{mylinkcolor}{violet}
\colorlet{mycitecolor}{violet}
\colorlet{myurlcolor}{violet}

\hypersetup{
  linkcolor  = mylinkcolor!\myshade!black,
  citecolor  = mycitecolor!\myshade!black,
  urlcolor   = myurlcolor!\myshade!black,
  colorlinks = true
}
\newcommand{\U}{\mathrm{U}}
\newcommand{\SU}{\mathrm{SU}}

\newcommand{\LL}{\mathrm{L}}
\newcommand{\RR}{\mathrm{R}}

\newcommand{\eminus}{\vcenter{\hbox{\scalebox{0.6}[1]{$ - $}}}}	
\newcommand{\rep}[1]{\mathbf{#1}}
\newcommand{\repbar}[1]{\overline{\mathbf{#1}}}

\newcommand{\sscript}[1]{{\scriptscriptstyle \mathrm{#1}}}

\def\L{\mathcal{L}}

\keywords{}

\begin{document}


\title{
 \boldmath $\mathrm{U(2)}$ is Right for Leptons and Left for Quarks
}
\author{Stefan Antusch}
\email{stefan.antusch@unibas.ch}
\affiliation{Department of Physics, University of Basel, Klingelbergstrasse 82,  CH-4056 Basel, 
Switzerland}

\author{Admir Greljo}
\email{admir.greljo@unibas.ch}
\affiliation{Department of Physics, University of Basel, Klingelbergstrasse 82,  CH-4056 Basel, 
Switzerland}

\author{Ben A. Stefanek}
\email{benjamin.stefanek@kcl.ac.uk}
\affiliation{Physics Department, King’s College London, Strand, London, WC2R 2LS, United Kingdom}

\author{Anders Eller Thomsen}
\email{thomsen@itp.unibe.ch}
\affiliation{Albert Einstein Center for Fundamental Physics, Institute for Theoretical Physics, University of Bern, CH-3012 Bern, Switzerland}


\preprint{KCL-PH-TH/2023-64}

\begin{abstract}
We posit that the distinct patterns observed in fermion masses and mixings are due to a minimally broken $\U(2)_{q+e}$ flavor symmetry acting on left-handed quarks and right-handed charged leptons, giving rise to an accidental $\U(2)^5$ symmetry at the renormalizable level without imposing selection rules on the Weinberg operator. We show that the symmetry can be consistently gauged by explicit examples and comment on realizations in $\SU(5)$ unification. Via a model-independent SMEFT analysis, we find that selection rules due to $\U(2)_{q+e}$ enhance the importance of charged lepton flavor violation as a probe, where significant experimental progress is expected in the near future.
\end{abstract}

\maketitle

\section{Introduction} 
\label{sec:intro}

The pattern of quark and lepton masses and their mixings under the weak nuclear force is peculiar: The masses of up quarks, down quarks, and charged leptons exhibit a generational hierarchy, and the CKM matrix~\cite{Kobayashi:1973fv} is conspicuously close to the unit matrix, showcasing a suppressed inter-generational mixing. On the contrary, neutrinos exhibit totally different behavior, as the PMNS matrix~\cite{Maki:1962mu} establishes significant mixing. The complete spectrum of neutrino masses remains elusive due to the unmeasured absolute mass scale ($\lesssim 10^{-6}\,m_e$) and mass ordering, leading to ambiguities between hierarchical and quasi-degenerate (or anarchic) scenarios. 

Within the context of the Standard Model (SM) viewed as an effective field theory (SMEFT), there exists a compelling explanation for the stark difference between the masses of neutrinos and charged fermions: neutrino masses arise from a higher-dimensional Weinberg operator~\cite{Weinberg:1979sa}. Yet, the SM fails to explain the observed hierarchy in charged fermion masses and the relative order in the CKM versus the apparent anarchy in the PMNS.

The spectrum of charged fermions can be traced back to their renormalizable interaction with a single Higgs field. These interactions are characterized by three distinct $3\times 3$ Yukawa matrices: $Y_u$, $Y_d$, and $Y_e$.  At the heart of the flavor puzzle lies the observation that the eigenvalues of these matrices manifest differences spanning two orders of magnitude between consecutive generations. This distinct behavior is consistently reflected across all three sectors, $u$, $d$, and $e$. Moreover, a striking feature is the approximate alignment observed between $Y_u$ and $Y_d$, as they are ostensibly independent matrices within the SM. These consistent patterns and alignments strongly suggest the need for an explanatory framework beyond the SM.

Work on longstanding puzzles in flavor physics is experiencing a resurgence, invigorated by an extensive, ongoing experimental program set to reach its zenith this decade. This renewed attention stems from the potential revelation of indirect effects of heavy new physics (NP) originating from scales well beyond the direct accessibility of the LHC, which has already reached its designed collision energy. Such effects could manifest in phenomena like rare flavor-changing neutral currents (FCNC), charged lepton flavor violation (cLFV), and electric dipole moments (EDMs), all of which will experience considerable experimental advancements~\cite{Belle-II:2018jsg, Forti:2022mti, Belle-II:2022cgf, LHCb:2018roe, LHCb:2021glh, HIKE:2022qra, NA62KLEVER:2022nea, MEGII:2018kmf, Bernstein:2019fyh, Moritsu:2022lem, n2EDM:2021yah, Wu:2019jxj, EuropeanStrategyforParticlePhysicsPreparatoryGroup:2019qin,Blondel:2013ia,ACME:2018yjb}. 
Selection rules will likely govern the observed flavor patterns and, if so, will determine the NP effects these experiments might detect. 
Here, we identify a potential origin of flavor hierarchies and study the correlated pattern of deviations in precision measurements due to indirect short-distance effects captured by the SMEFT.

The SM Yukawa couplings contain limited information about the violation of the full $ \U(3)^5 $ flavor symmetry of the fermion kinetic terms.
We adopt the perspective that the observed approximate symmetries may be largely accidental. Specifically, we impose a non-Abelian flavor symmetry group $\U(2)_{q+e}$ that realizes the distinct flavor patterns observed in the quark and lepton sectors.\footnote{To our knowledge, this $\U(2)$ flavor symmetry was not considered in existing literature~\cite{Barbieri:1995uv, Barbieri:1996ae, Barbieri:1996ww, Barbieri:1999pe, Masiero:2001cc, Linster:2018avp, Barbieri:1997tu, Carone:1997qg,  Roberts:2001zy, Raby:2003ay, Dudas:2013pja, Falkowski:2015zwa, Feruglio:2019ybq, Barbieri:2019zdz, Linster:2020fww, Greljo:2023bix}.}
By enforcing this symmetry on the SM viewed as an EFT, the renormalizable Lagrangian inadvertently respects a larger accidental symmetry, $\U(2)^5$~\cite{Feldmann:2008ja, Kagan:2009bn, Barbieri:2011ci, Barbieri:2012uh, Blankenburg:2012nx, Fuentes-Martin:2019mun, Faroughy:2020ina, Greljo:2022cah}. 
The symmetry, therefore, predicts large, unsuppressed masses for the third generation of all charged fermions, while the light generations remain essentially massless. 

However, only $\U(2)_{q+e}$ is respected beyond the renormalizable level, so the setup addresses flavor hierarchies in charged fermions without imprinting the same into the neutrino sector. The crucial point is that $\U(2)_{q+e}$ does not enforce selection rules on the Weinberg operator, suggesting an anarchic PMNS matrix. The ultraviolet (UV) origins of this symmetry will be explored in Section~\ref{sec:model}.

The minimal breaking of $\U(2)_{q+e} \to \U(1)_{q_1+e_1}$ via a single spurion doublet predicts a massive second and massless first generation, explaining the hierarchy between them. Including a subdominant second spurion doublet ensures that the first generation is not left massless. As a result, not only are the masses between generations hierarchical, but left-handed quark mixings also follow the ratio of masses, which produces the observed CKM matrix. Our setup also justifies why the hierarchy between generations is common to up- and down-type quarks, as well as charged leptons.

\section{Symmetry behind fermion masses and mixings} 
\label{sec:spurion}
We charge left-handed quarks and right-handed charged leptons under a global $\U(2)_{q+e} \equiv \SU(2)_{q+e} \times \U(1)_{q+e}$ symmetry. The flavor triplets decompose as $ q^p = q^\alpha \oplus q^3 \sim \rep{2}_1 \oplus \rep{1}_0 $ and $ e^p = e^\alpha \oplus e^3 \sim \rep{2}_1 \oplus \rep{1}_0 $, whereas all other SM fields are singlets under the new symmetry. Here $\alpha=1,2$ designates the $\SU(2)_{q+e}$ index. In the minimal scenario, $ \U(2)_{q+e} $ is broken by two spurions $ V^\alpha_{1,2} \sim \rep{2}_1 $. We identify $ V^\alpha_2 = (0,\, a) $ and $ V^\alpha_1 = (b,\, 0) $.\footnote{A $ \U(2)_{q+e} $ rotation guarantees the alignment of $ V_2 $, while the part of $ V_1 $ parallel to $ V_2 $ can be neglected given the size hierarchies.} Generation hierarchies are explained with the assumption $ 1 \gg a \gg b >0  $.

The symmetric Yukawa couplings are  
    \begin{align} \label{eq:Yukawas}
    \L \supset &- \big( x_u^p \overline{q}^3 + y_u^p \overline{q}_\alpha V_2^\alpha + z_u^p \overline{q}_\alpha V_1^\alpha \big) \widetilde H u^p \nonumber \\
    &- \big( x_d^p \overline{q}^3 + y_d^p \overline{q}_\alpha V_2^\alpha  + z_d^p \overline{q}_\alpha V_1^\alpha \big) H d^p \\
    &- \overline{\ell}^p H \big( x_e^p e^3 + y_e^p V^\ast_{2\alpha} e^\alpha  + z_e^p V^\ast_{1\alpha} e^\alpha \big)\; + \mathrm{H.c.} \nonumber
    \end{align}
including insertions of the spurions. The couplings $ x_f^p, y_f^p, z_f^p$ with $ f=\{u,\, d,\, \ell\} $ and $ p = \{1,\, 2,\, 3\}$ are expected to be of the same order, say $\mathcal{O}(0.3)$. Using $ \U(3)_u \times \U(3)_d \times \U(3)_\ell $ transformations to remove unphysical parameters, the SM Yukawa matrices can be put into a triangular form, e.g.,
    \begin{equation} \label{eq:Yu_matrix}
    Y_u = \begin{pmatrix}
        z_{u1} b & z_{u2} b& z_{u3} b \\
        & y_{u2} a& y_{u3} a \\ & & x_{u3} 
    \end{pmatrix} = L_u \widehat{Y}_u R_u^\dagger  
    \end{equation}
for the up-type quarks. The products of singular value decompositions are the positive diagonal matrices $ \widehat{Y}_f $ and unitary rotation matrices $ L_f, R_f$.

The triangular form of the Yukawa matrices with hierarchies between the rows (columns) for quarks (charged leptons) admits perturbative diagonalization as demonstrated in App.~\ref{app:perturbative_diagonalization}. The Higgs couplings to the mass eigenstates ($ \widehat{Y}_f $) are given by the diagonal values of the coupling matrices ($ Y_f$) to the leading order. Meanwhile, the expected sizes of the mixing angles are determined by the corresponding mass ratios. The down quark mass hierarchy is compressed compared with the up-type quarks on account of the smallness of the bottom quark mass. 
Thus, the resulting CKM matrix, $ L^\dagger_u L_d $, is expected to be dominated by the down-type rotation. It is approximately unity, while the size of the off-diagonal elements agrees well with the ratio of down quark masses.
We show a numerical benchmark in App.~\ref{app:perturbative_diagonalization} that demonstrates that generation hierarchies can be explained by the hierarchical spurions.\footnote{The largest departures from $ \mathcal{O}(0.3) $ couplings are the bottom and tau Yukawas $ x_{d3}, x_{\ell3} \simeq 0.01 $. See Section~\ref{sec:GUT} for an elegant explanation.} 

We can now appreciate why a minimally broken $ \U(2)_{q+e} $ economically explains the observed flavor patterns. A $ \U(2)_q $ symmetry is sufficient to produce the hierarchies of the quark masses while ensuring a near-identity CKM matrix. 
On the other hand, a $ \U(2)_\ell $ symmetry is unsuited for the lepton sector: in analogy with the quark symmetry, it would tend to generate a near-identity PMNS matrix in contradiction with observation. Instead, $ \U(2)_e $ stands out as a good explanation of the charged lepton mass hierarchies without restricting the PMNS mixing. Finally, the quark and lepton hierarchies being similar lends itself to the identification of the two symmetries: $ \U(2)_q = \U(2)_e $. In this respect, $ \U(2)_{q+e} $ represents the minimal choice; however, additionally, charging up-type quarks can be beneficial (cf. Section~\ref{sec:GUT}).

Readers might wonder about the $ \U(1) $ factor. In its absence, there would exist Yukawa couplings to both $ V^\alpha_i $ and $ \widetilde V^\alpha_i = \varepsilon^{\alpha \beta} V^\ast_{i \alpha} $, which would wash out the hierarchy between the first and second generation fermions. Some $ \U(1) $ factor is required to explain all the hierarchies; however, there are many possible variations.

Let us reiterate that no selection rules are imposed on the Weinberg operator, $\ell^p \ell^r H H$. Here, the dimension-5 Weinberg operator, or higher-dimensional Weinberg-like operators, serve as a placeholder for a neutrino mass generation, and we do not need to commit to a particular model. The essence lies in the absence of selection rules for left-handed leptons, starkly contrasting the restrictions present when applying $\U(2)_\ell$~\cite{Greljo:2023bix}. Here, we instead elegantly reconcile charged fermions with neutrinos.

\begin{figure*}
  \centering
  \includegraphics[trim={0.2cm 0cm .95cm 0cm}, clip, width=1\textwidth]{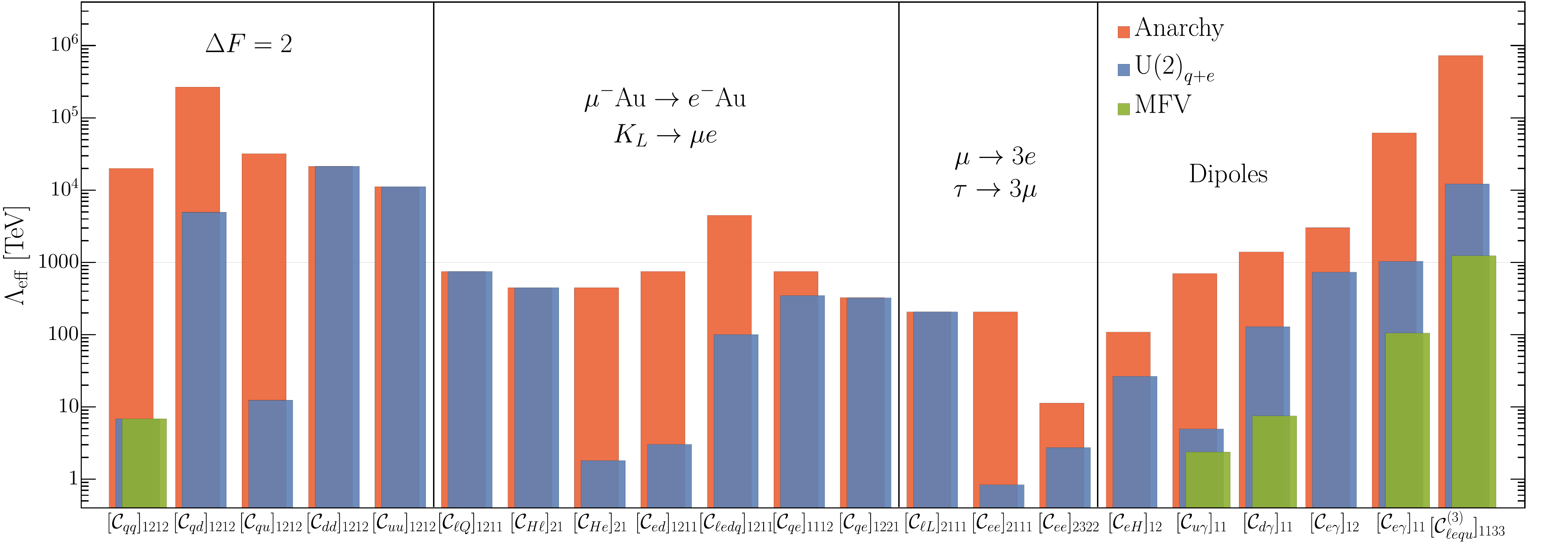}
  \caption{Comparative constraints on SMEFT operators from flavor and CP violation: Minimally-broken $\U(2)_{q+e}$ (Blue), MFV (Green), Flavor Anarchy (Red). Here, $Q=q,u,d$ and $L = \ell,e$. See Section~\ref{sec:SMEFT} for details.}
  \label{fig:bounds}
\end{figure*}

\section{$\U(2)_{q+e}$ selection rules in the SMEFT} 
\label{sec:SMEFT}

While the $ \U(2)_{q+e} $ symmetry is indistinguishable from $ \U(2)^5 $ in the renormalizable SM, it produces vastly different selection rules for UV physics. We use the SMEFT as a proxy for high-scale NP to study the general implications of the symmetry. Imposing $ \U(2)_{q+e} $ on the SMEFT organizes a power counting in the spurion insertions. 

Excluding baryon number violation, the largest scales are probed by flavor- and CP-violating observables. We have identified four different sources of flavor violation in our setup: 
1) symmetry-unprotected operators involving right-handed quarks and left-handed leptons, which are singlets under $ \U(2)_{q+e} $; 
2) flavor-transferring operators~\cite{Darme:2023nsy} where the $ \U(2)_{q+e} $ quantum numbers flow from $ q^\alpha $ to $e^\alpha $ without extra suppression; 
3) flavor violation from the perturbative rotations that bring the SM fermions from the interaction to the mass basis; 
and 4) direct insertions of the spurions $ V_{1,2}$ in the SMEFT. 

We expect that the rotations dominate over direct spurion insertions for observables involving left-handed down quarks and right-handed charged leptons. As previously mentioned, the smallness of $ y_b $ and $ y_\tau $ enhances the size of the rotations, which we would otherwise expect to be of the same size as direct spurion insertions. In our numerical analysis, neglecting $\mathcal{O}(1)$ constants, we consider the perturbative left-handed down quark rotation $L_d = V_{\rm CKM}$, left-handed up quark rotation 
$L_u^{\{12,13,23\}} = \{b/a,b,a\}$, and right-handed charged lepton rotation $R_e^{ij} = m_{e_i}/m_{e_j}$ ($i \leq j$) applied to $\U(2)_{q+e}$--symmetric operators, all evaluated at the scale $\mu = 1$\,PeV.\footnote{For the left-handed up-type quarks, direct spurion insertions are comparable in contribution to perturbative rotations.} 

Without specifying a UV completion, we systematically examine each operator within the Warsaw basis one by one~\cite{Grzadkowski:2010es}, deriving the present experimental constraints by adopting the $\U(2)_{q+e}$ flavor structure. We report limits on the effective scale $\Lambda_{\text{eff}}$, $\mathcal{L} \supset \frac{1}{\Lambda^2_{\text{eff}}} \mathcal{O}$, ensuring that the operator $\mathcal{O}$ includes the relevant flavor suppression.  We also incorporate a loop suppression factor for dipole operators, as suggested by perturbative UV completions. The procedure similarly applies to MFV~\cite{DAmbrosio:2002vsn} and flavor anarchy scenarios to provide comparative benchmarks.

The most sensitive indirect experimental probes can be organized into the categories $\Delta F = 2$~\cite{Silvestrini:2018dos}, 
cLFV~\cite{Feruglio:2015gka, Davidson:2020hkf}, and EDMs~\cite{ACME:2018yjb, Kley:2021yhn, Marzocca:2021miv}, and we collect the reported limits from these references. The resulting limits on the effective scales for various operators are shown in~\cref{fig:bounds}. Operators are specified by a Wilson coefficient, adopting the notation of Ref.~\cite{Grzadkowski:2010es}, with the flavor indices that exhibit the strongest bounds.

Our results can be succinctly summarized as follows: First, operators involving only the fields $\ell$, $u$, $d$, or flavor transfers of the form $\bar q_\alpha e^\alpha$ are not suppressed by $\U(2)_{q+e}$ and, therefore, exhibit the same bounds as the flavor anarchy scenario. The same is true for flavor-violating $\ell$, $u$, or $d$ currents that multiply flavor-conserving currents of any species, e.g., $[\mathcal{C}_{\ell q}]_{1211}$ contributing to $\mu\rightarrow e$ conversions or $[C_{\ell e}]_{2111}$ producing $\mu \rightarrow 3 e$. 
On the other hand, flavor-violating $q$-currents are CKM suppressed, resulting in MFV-like bounds for $4q$ operators and moderate suppression in mixed $2q$ operators. Similarly, flavor-violating $e$-currents are $R_e$ suppressed, for example, $\mu \to 3 e$ from $\mathcal{O}_{ee}$ arises only via $[R_e^\dag]_{23} [R_e]_{31} \propto m_e m_\mu /m_\tau^2$. 

Chirality-flipping operators built from $\bar\ell e$, $\bar q u$, or $\bar q d$ break $\U(2)_{q+e}$ and come suppressed compared to anarchy. Examples include dipole operators, as well as scalar and tensor 4-fermion operators. Especially, the small left-handed up quark rotation gives a strong suppression in $\bar q u$ combinations. An interesting case is the $[\mathcal{O}^{(3)}_{lequ}]_{1133}$ operator, which gives large contributions to electron EDM via RG mixing involving the top Yukawa~\cite{Jenkins:2013wua}.

The SMEFT analysis provides a useful guideline for the expected pattern of deviations. 
Fig.~\ref{fig:bounds} illustrates that $\U(2)_{q+e} $ gives large suppression for many operators compared with anarchy. In particular, this is the case for the strongest bounds from the electron EDM, $\mu \rightarrow e\gamma$, and $ \Delta F=2 $, so we may expect a lower NP scale than in anarchic models.\footnote{The NP scale is $\gtrsim$ PeV due to unsuppressed cLFV and right-handed meson mixing. To allow for TeV-scale NP, the SMEFT should respect a larger flavor symmetry~\cite{DAmbrosio:2002vsn, Greljo:2023adz, Allwicher:2023shc}.} The suppression puts these bounds on the same footing as those from other cLFV processes such as $\mu\rightarrow e$ conversion, $K_L \rightarrow \mu e$, and $\mu \rightarrow 3e$ that are largely unsuppressed and are therefore more competitive in our scenario. This is especially interesting in light of the expected future sensitivity of the Mu3e~\cite{Blondel:2013ia}, Mu2e~\cite{Bernstein:2019fyh}, COMET~\cite{Moritsu:2022lem}, and ACME~\cite{ACME:2018yjb, Wu:2019jxj} experiments, which will push the sensitivity of $\mu \rightarrow 3e$, $\mu\rightarrow e$ conversions, and the electron EDM to NP by order of magnitude in the effective scale relative to current bounds.

\section{Gauged $\SU(2)_{q+e}$ completions} 
\label{sec:model}

When looking for renormalizable UV explanations of the $\U(2)_{q+e}$ symmetry, we are drawn to consider gauging the non-Abelian component. Thus, we extend the SM gauge symmetry to $\SU(3)_{\text{c}}\times \SU(2)_\LL \times \U(1)_\text{Y}\times \SU(2)_{q+e}$,
where the fermions are charged under $\SU(2)_{q+e}$ as described in Sec.~\ref{sec:spurion}.
Additional chiral fermions are needed to ensure anomaly cancellation. An expedient, but not unique, choice of anomalons\footnote{Local anomalies notably arise from triangular diagrams associated with $\U(1)_Y \times \SU(2)_{q+e}^2$. For example, an alternative to a right-handed doublet with $Y=2$ would be a color triplet weak singlet with $Y=2/3$, etc. Note that the global $\SU(2)_{q+e}$ anomaly mandates an even total number of doublets~\cite{Witten:1982fp, Davighi:2019rcd}. The additional left-handed states are $\SU(2)_{q+e}$ singlets, ensuring the new states appear vector-like under the SM gauge group. 
} features a right-handed $\Psi^\alpha_\RR \sim (\rep{1}, \rep{1},+2,\rep{2})$ and two copies of left-handed $\Psi^i_\LL \sim (\rep{1}, \rep{1},+2,\rep{1})$, with $i=1,2$ marking a flavor index. A scalar field $\Phi^\alpha \sim (\rep{1}, \rep{1},0,\rep{2})$ induces spontaneous breaking of the $\SU(2)_{q+e}$ symmetry via its VEV, $|\langle \Phi \rangle | = v_\Phi$. In scenarios where $v_\Phi \gg v_{\text{EW}}$, the associated gauge bosons become sufficiently massive, skirting experimental detection, as detailed subsequently. Likewise, the anomalons become massive due to $\mathcal{L} \supset -\eta_i \overline{ \Psi}_\RR \Phi \Psi^i_\LL - \tilde \eta_i  \overline{\Psi}_\RR \widetilde \Phi \psi^i_\LL$, where $\widetilde \Phi \equiv \epsilon \Phi^*$.\footnote{This Lagrangian conserves $\U(1)_\Psi$ number, predicting that the lightest state is stable. However, higher-dimensional operators like $(\overline{\Psi}_\LL \gamma_\mu \tau^C_\RR)(\overline{d}_\RR \gamma^\mu u_\RR)$ can disrupt this, if demanded by viable cosmology.}

The selection rules implied by the extended gauge symmetry allow for Higgs Yukawa interactions only with $q^3$ and $e^3$, predicting massive third and massless light generations of charged fermions. Introducing a single copy of vector-like fermions $Q_{\LL,\RR}\sim (\rep{3}, \rep{2},\frac{1}{6},\rep{1})$ and $E_{\LL,\RR}\sim (\rep{1},\rep{1},-1,\rep{1})$ with masses $M_Q$ and $M_E$, respectively, allows for additional Higgs interactions $\mathcal{L} \supset -y_u^p \bar Q \tilde H u^p - y_d^p \bar Q H d^p - y_e^p \bar \ell^p H E$, where $p=1,2,3$. Other important renormalizable interactions are with the $\Phi$ field, $\mathcal{L} \supset -\kappa_q \bar q \Phi Q - \tilde \kappa_q \bar q \widetilde \Phi Q - \kappa_e \bar e \Phi E - \tilde \kappa_e \bar e \tilde \Phi E$. Integrating out heavy vector-like fermions at tree level predicts effective Yukawa matrices of rank~2, giving a mass only to the second family, leading to an accidental $U(1)^5$ symmetry for the first family matter fields. 
Assuming for simplicity that $\kappa_q / \tilde \kappa_q = \kappa_e / \tilde \kappa_e$,\footnote{When this is not the case, there is an additional $\mathcal{O}(1)$ rotation in the 1--2 sector of the right-handed charged leptons.} and $M_Q = M_e$, the spurion $V_2^\alpha$ introduced in Section~\ref{sec:spurion} is identified with
\begin{equation}
    V_2^\alpha = \frac{\kappa_q}{M_Q} \langle \Phi^\alpha \rangle + \frac{\tilde\kappa_q}{M_Q} \langle \widetilde \Phi^\alpha \rangle~.
\end{equation}
Numerically, $M_Q = M_E \approx 100 \,v_\Phi$ explains the 2-3 hierarchy. 

Finally, there are multiple options for generating mass for the first family. Trivially, introducing heavier copies of $Q$ and $E$ would induce a suppressed $V^\alpha_1$ spurion. Alternatively, a set of heavy scalars $\Phi'$ and $H'$ (same quantum numbers as $\Phi$ and $H$) that do not get VEVs can radiatively generate masses for the first family. In particular, the new scalars run in the loop together with the vector-like fermions $Q$ or $E$ already present for the second family. This is facilitated by quartic interactions such as $\Phi^\dagger \Phi' H^\dagger H' $. The resulting effective Higgs Yukawas for the first family will be suppressed by a loop factor with respect to those for the second family. The SM flavor structure is rather robust since it is largely insensitive to the spectrum of heavy scalars $\Phi'$ and $H'$ as long as they are lighter than $Q,E$ (or comparable). An alternative UV completion could utilize anomalons in the loop to produce dim-5 Yukawa operators for both the first and second generation, removing the need for vector-like fermions. This path would require extra scalar leptoquarks running in the loop. Various options call for future exploration of different UV completions.

In all cases, the irreducible phenomenological implication of a gauged model is the tree-level exchange of $Z'$ bosons, resulting in the generation of $\mathcal{O}_{qq}$, $\mathcal{O}_{ee}$, and $\mathcal{O}_{qe}$ operators. Although $\mathcal{O}_{qq}$ and $\mathcal{O}_{ee}$ are notably suppressed, as depicted in Fig.~\ref{fig:bounds}, $\mathcal{O}_{qe}$ induces substantial $\mu \to e$ conversion and the decay $K_L \to \mu e$. This imposes a current lower bound on the symmetry-breaking scale of $v_\Phi \gtrsim$\,PeV~\cite{Greljo:2023bix,Darme:2023nsy}. As previously mentioned, significant improvement is expected in the NP sensitivity of $\mu \rightarrow e$ conversions by the MU2e and COMET experiments, making them the leading probe of such models with a reach of up to 10 PeV.

\section{A GUT-inspired variant}
\label{sec:GUT}

Another global flavor symmetry closely related to $\text{U}(2)_{q+e}$ has an attractive realization in $ \SU(5) $ grand unified theories (GUTs)~\cite{Georgi:1974sy}. Let only the first two generations of the matter representation $\rep{10}^\alpha$ ($\alpha=1,2$) form a doublet of $ \U(2) $. Then the right-handed up-type quarks are embedded along with $q$ and $e$, extending the flavor symmetry to $ \U(2)_{10} \equiv \U(2)_{q+e^c+u^c} $. All lepton doublets and right-handed down-type quarks, contained in the $\repbar{5} $'s, are singlets under $ \U(2)_{10}  $, explaining the anarchic PMNS mixing compatible with the hierarchical charged fermion masses and the CKM mixing, as in Section~\ref{sec:spurion}. This elegant option was not explored in the previous literature on $ \U(2) $ flavor symmetries in GUTs~\cite{Barbieri:1996ae, Barbieri:1996ww, Barbieri:1999pe, Masiero:2001cc, Linster:2018avp}, which, by charging left-handed leptons, introduces unwanted hierarchies in the neutrino sector that has to be corrected with an additional structure.

With the $\rep{10}^\alpha$ forming doublets under $ \U(2)_{10} $, the up- and charm-quark Yukawa couplings (from $V_{i\alpha} V_{j\beta}  \rep{10}^\alpha \rep{10}^\beta  \rep{5}_H$) are doubly suppressed by spurion insertions, whereas the top Yukawa coupling is unsuppressed. On the contrary, the light family down-type quark and charged lepton Yukawas ($V_{i\alpha} \repbar{5} \,\rep{10}^\alpha \rep{5}^*_H$) receive only a single spurion suppression, effectively explaining the compressed hierarchy compared with the up-quark sector. The predicted structure of the left-handed rotation matrices is now similar for the up- and down-type quarks. At this point, the only unexplained structure is the overall normalization of the down-type quark and charged lepton masses. This is naturally explained by an additional symmetry acting universally on the $\repbar{5}^p$ ($p=1,2,3$). The simplest example is presumably a $\mathbb{Z}_2$ symmetry where the $\repbar{5}^p$ and a spurion $V_Z$ are odd, while all other fields are even. A plausible dynamical realisation is a type-II two-Higgs doublet model with hierarchical VEVs, i.e.\  large $\tan \beta$. Remarkably, this scenario produces all fermion masses and mixings from $\mathcal{O}(1)$ interactions, including the smallness of $y_b$ and $y_\tau$.\footnote{The order-of-magnitude predictions from $ \U(2)_{10} \times \,\mathbb{Z}_2$ are $y_t \sim 1$, $y_b\sim y_\tau \sim V_Z$, $\frac{y_s}{y_b} \sim \frac{y_\mu}{y_\tau} \sim |V_{cb}| \sim \sqrt{y_c} \sim a$, $\frac{y_d}{y_b} \sim \frac{y_e}{y_\tau} \sim | V_{ub}| \sim \sqrt{y_u} \sim b$, while $|V_{us}| \sim \frac{b}{a}$. All observed parameters are within their anticipated order of magnitude for $ V_Z = 0.01$, $a  = 0.03 $ and $b = 0.002$. See App.~\ref{app:perturbative_diagonalization} for details.}

The analogous SMEFT analysis differs from the one reported in Section~\ref{sec:SMEFT}, particularly for the operators involving up-type quarks. In addition, we note that $\SU(2)_{10}$ can be gauged without introducing chiral anomalies, leaving the realistic model construction for future work.   


\section{Outlook and Conclusions} 
\label{sec:conc}

With an eye to the future, taking $\U(2)_{q+e}$ (or $\U(2)_{q+e^c+u^c}$) flavor symmetry as a new starting point opens up various unexplored routes. As discussed above, this involves UV completions where the symmetry is gauged, investigating its variants with different $\U(1)$ factors and extending to the next layer, e.g., in the context of GUTs. Furthermore, its implications for SMEFT analyses with flavor symmetries, predictions for soft supersymmetry breaking terms, proton decay, etc., will be interesting to study. We have illustrated some of the initial ideas in Sections~\ref{sec:SMEFT} and \ref{sec:model}.  



In closing, we have shown that the distinct patterns observed in fermion masses and mixings could point towards a unifying $\U(2)_{q+e}$ (or $\U(2)_{q+e^c+u^c}$) flavor symmetry, from which the approximate $\U(2)^5$ symmetry in the SM Yukawa couplings emerges accidentally. By charging light-family left-handed quarks and right-handed leptons under a minimally-broken $\U(2)$, we obtain an explanation for the generation hierarchies in charged fermion masses and the CKM mixing while simultaneously imposing no selection rules on the Weinberg operator. The latter feature allows for small neutrino masses and predicts large (anarchic) mixing. As shown in Fig.~\ref{fig:bounds}, our symmetry suppresses NP contributions to the most sensitive SMEFT operators, making certain cLFV processes more competitive.
Intriguingly, improved measurements of $\mu \rightarrow 3e$, $\mu\rightarrow e$ conversions, and the electron EDM planned in the near future will push the sensitivity to the NP scale by an order of magnitude. Hopefully, new clues will illuminate a path forward, casting light on our quest to unravel the origins of flavor.

\section*{Acknowledgments}

This work has received funding from the Swiss National Science Foundation (SNF) through the Eccellenza Professorial Fellowship ``Flavor Physics at the High Energy Frontier,'' project number 186866, and the Ambizione grant ``Matching and Running: Improved Precision in the Hunt for New Physics,'' project number 209042.

\appendix 
\renewcommand{\thesection}{\Alph{section}}
\renewcommand{\thesubsection}{\Alph{section}.\arabic{subsection}}
\setcounter{section}{0}

\section{Perturbative diagonalization} \label{app:perturbative_diagonalization}

\paragraph{$U(2)_{q+e}$} Here we provide the essential part of the perturbative diagonalization of the Yukawa couplings and provide a benchmark parameter point. Similarly to the up-Yukawa matrix~\eqref{eq:Yu_matrix}, the 
$ \U(3)_u \times \U(3)_d \times \U(3)_\ell $ transformations allows us to simultaneously put the down quark and charged lepton Yukawa matrices in triangular form: 
    \begin{align} \label{eq:Yd_Ye_matrix}
    Y_d &= \begin{pmatrix}
        z_{d1} b & z_{d2} b& z_{d3} b \\
        & y_{d2} a& y_{d3} a \\ & & x_{d3} 
    \end{pmatrix} = L_d \widehat{Y}_d R_d^\dagger,\\
    Y_e &= \begin{pmatrix}
        z_{\ell 1} b& & \\
        z_{\ell 2} b& y_{\ell 2}a & \\
        z_{\ell 3} b & y_{\ell 3} a& x_{\ell 3}
    \end{pmatrix}= L_e \widehat{Y}_e R_e^\dagger.
    \end{align}
The singular value decomposition of all three Yukawa matrices can be determined perturbatively when $ b \ll a\ll 1$. The diagonal elements of the Yukawa matrices are the leading contribution to the corresponding singular values, meaning that, e.g., $ m_s/v_\mathrm{EW} \simeq y_{d2} a $. 
The rotation matrices are given by 
    \begin{align}
    L_d \simeq \begin{pmatrix}
            1 & \tfrac{m_d}{m_s} \tfrac{z_{d2}}{z_{d1}} & 
            \tfrac{m_d}{m_b} \tfrac{z_{d3}}{z_{d1}} \\
            -\tfrac{m_d}{m_s} \tfrac{z^\ast_{d2}}{z_{d1}} & 1 &
            \tfrac{m_s}{m_b} \tfrac{y_{d3}}{y_{d2}} \\ 
            \tfrac{m_d}{m_b} \big(\tfrac{y^\ast_{d3} z^\ast_{d2}}{y_{d2} z_{d1}} - \tfrac{z_{d3}^\ast}{z_{d1}} \big) &
            - \tfrac{m_s}{m_b} \tfrac{y^\ast_{d3}}{y_{d2}} & 1
        \end{pmatrix}, \\
    L_u \simeq \begin{pmatrix}
            1 & \tfrac{m_u}{m_c} \tfrac{z_{u2}}{z_{u1}} & 
            \tfrac{m_u}{m_t} \tfrac{z_{u3}}{z_{u1}} \\
            -\tfrac{m_u}{m_c} \tfrac{z^\ast_{u2}}{z_{u1}} & 1 &
            \tfrac{m_c}{m_t} \tfrac{y_{u3}}{y_{u2}} \\ 
            \tfrac{m_u}{m_t} \big(\tfrac{y^\ast_{u3} z^\ast_{u2}}{y_{u2} z_{u1}} - \tfrac{z_{u3}^\ast}{z_{u1}} \big) &
            - \tfrac{m_c}{m_t} \tfrac{y^\ast_{u3}}{y_{u2}} & 1
        \end{pmatrix}, \\
    R_e \simeq \begin{pmatrix}
            1 & \tfrac{m_e}{m_\mu} \tfrac{z^\ast_{\ell2}}{z_{\ell1}} & 
            \tfrac{m_e}{m_\tau} \tfrac{z^\ast_{\ell3}}{z_{\ell1}} \\
            -\tfrac{m_e}{m_\mu} \tfrac{z_{\ell2}}{z_{\ell1}} & 1 &
            \tfrac{m_\mu}{m_\tau} \tfrac{y^\ast_{\ell3}}{y_{\ell2}} \\ 
            \tfrac{m_e}{m_\tau} \big(\tfrac{y_{\ell3} z_{\ell2}}{y_{\ell2} z_{\ell1}} - \tfrac{z_{\ell3}}{z_{\ell1}} \big) &
            - \tfrac{m_\mu}{m_\tau} \tfrac{y^\ast_{\ell3}}{y_{\ell2}} & 1
        \end{pmatrix},
    \end{align}
where we have included the leading contribution in each entry. The masses of the SM fermions are the running masses at the matching scale to the SM. 

As a realistic benchmark point for the $ \U(2)_{q+e} $ symmetry, we consider a SMEFT matching scale of $ \mu = \SI{1}{PeV} $. At this scale, we take~\eqref{eq:Yukawas} to reproduce the SM parameters~\cite{Martin:2019lqd} run up to the matching scale~\cite{Thomsen:2021ncy}. 
We take the spurions $ (a,\, b) = (3\cdot 10^{\eminus 3},\, 5\cdot 10^{\eminus 5})$, which allows us to determine the parameters  
    \begin{align}
        z_{\ell1} & = 0.057 & y_{\ell2} & = 0.20 & x_{\ell3} & = 0.010 \nonumber\\
        z_{u1} & = 0.091 & y_{u2} & = 0.76 & x_{u3} & = 0.67 \\
        z_{d1} & = 0.20 & y_{d2} & = 0.066 & x_{d3} & = 0.010 \nonumber\\
        z_{d2} & = 0.89e^{i\alpha} & z_{d3} & = 0.72e^{i(\beta -1.2)} & y_{d3} & = 0.13 e^{i(\beta-\alpha)} \nonumber 
    \end{align}
with arbitrary phases $ \alpha, \beta$, under the assumption that $ L_d = V_\sscript{CKM}$. All of the couplings are roughly within an order of magnitude save for the bottom and tau Yukawas.

\paragraph{$ \U(2)_{q+e^c+u^c} \times \mathbb{Z}_2$} The down-quark and charged-lepton Yukawa matrices become
\begin{equation} \label{eq:Yd_Ye_matrix_10}
    Y_d = V_Z \! \begin{pmatrix}
        z_{d1} b & z_{d2} b& z_{d3} b \\
        & y_{d2} a& y_{d3} a \\ & & x_{d3} 
    \end{pmatrix},\;
    Y_e = V_Z \! \begin{pmatrix}
        z_{\ell 1} b& & \\
        z_{\ell 2} b& y_{\ell 2}a & \\
        z_{\ell 3} b & y_{\ell 3} a& x_{\ell 3}
    \end{pmatrix},
\end{equation}
while for the up-quark
\begin{equation} \label{eq:Yu_matrix_10}
    Y_u = \begin{pmatrix}
    z_{u1} b^2 & z_{u2} a b& z_{u3} b \\
       y_{u1} a b & y_{u2} a^2 & y_{u3} a \\ x_{u1} b & x_{u2} a & x_{u3} 
    \end{pmatrix}.
    \end{equation}
After perturbative diagonalisation, and setting the spurions $ (V_Z, a, b)=(0.01, 0.03, 0.002)$, we fit the flavor parameters with
\begin{align}
        z_{\ell1} & = 0.14 & y_{\ell2} & = 2.0 & x_{\ell3} & = 1.0 \nonumber\\
        z_{u1} & = 1.1 & y_{u2} & = 2.5 & x_{u3} & = 0.67 \\
        z_{d1} & = 0.50 & y_{d2} & = 0.66 & x_{d3} & = 1.0 \nonumber\\
        z_{d2} & = 2.2 e^{i\alpha} & z_{d3} & = 1.8e^{i(\beta -1.2)} & y_{d3} & = 1.3 e^{i(\beta-\alpha)} \nonumber 
\end{align}
where, for simplicity, we assumed the off-diagonal terms in $Y_u$ to be small. When those are of $\mathcal{O}(1)$, a comparable contribution to the up-quark masses and the CKM matrix is generated. Thus, $ x_f^p, y_f^p, z_f^p$ of $\mathcal{O}(1)$ correctly predict the observed flavor hierarchies.

\bibliographystyle{JHEP}
\bibliography{refs.bib}

\end{document}